# Giant and Rapidly Switching Intrinsic Chirality Enabled by Toroidal Quasi-Bound States in the Continuum

**Shijie Kang[1,2], Jiusi Yu[1], Boyuan Ge[1], Jiayu Fan[1], Aoning Luo[1], Yiyi Yao[1], Xiexuan Zhang[1], Ken Qin[1], Bo Hou[3], Haitao Li[1,*], Xiaoxiao Wu[1,2,*]**

[1]*Modern Matter Laboratory and Advanced Materials Thrust, The Hong Kong University of Science and Technology (Guangzhou), Nansha, Guangzhou 511400, China*

[2]*Low Altitude Systems and Economy Research Institute, The Hong Kong University of Science and Technology (Guangzhou), Nansha, Guangzhou 511400, Guangdong, China*

[3]*Wave Functional Metamaterial Research Facility (WFMRF), The Hong Kong University of Science and Technology (Guangzhou), Nansha, Guangzhou 511400, Guangdong, China*

**haitaoli@hkust-gz.edu.cn, xiaoxiaowu@hkust-gz.edu.cn*

**Abstract:** Circular dichroism (CD), arising from spin-selective light-matter interactions controlled by chirality, is critical for advanced applications such as chiral imaging and ultrasensitive biosensing. However, CD of chiral natural materials is inherently constrained owing to molecular symmetry and thermodynamic stability. Recently, artificially engineered metasurfaces incorporating chiral quasi-bound states in the continuum (Q-BICs) have emerged as a promising solution, which enables near-unity CD responses. However, their current designs heavily rely on complex three-dimensional geometries, posing significant challenges for integration with planar on-chip platforms. To address the stringent challenges, we demonstrate a truly planar metasurface that achieves giant intrinsic chiral responses by utilizing a chiral Q-BIC dominated by out-of-plane component ($T_z$) of toroidal dipole (TD). With deep-subwavelength (<λ/20) thickness, our metasurface exhibits outstanding intrinsic CD values in both simulations (>0.90) and experiments (~0.80). Moreover, in contrast to previous electric or magnetic chiral Q-BICs, the toroidal Q-BIC produces a rapidly switching CD response—transitioning sharply between positive and negative giant CD values within ~0.2 GHz, and the switching is highly sensitive to small oblique incidence of opposite angles. Therefore, our scheme provides a planar platform for studying chiral light-matter interactions involving TD, important for future development of polarization- and angle-sensitive photonic and optoelectronic devices.

## 1. Introduction

Chiral metamaterials have gained considerable attention in recent years owing to their exceptional capabilities in modulating light-matter interactions, enabling applications such as circular dichroism (CD) spectroscopy [1, 2], optical dispersion engineering [3, 4], and polarization-resolved imaging [5, 6, 7]. Their unique ability to discriminate handedness is fundamentally significant, not only because chirality is inherent in many natural and biological systems, determining molecular functionality and chemical selectivity, but also because it offers powerful means to tailor interactions at the nanoscale. These properties are critical for advances in enantioselective sensing [8, 9, 10], optical communication [11, 12, 13, 14, 15], and quantum optics [16, 17, 18, 19]. Early investigations predominantly focused on three-dimensional (3D) chiral architectures [20] that exploit spatial asymmetry to interact effectively with circularly polarized light (CPL). However, the complex fabrication processes, high production costs, and limited compatibility with planar photonic technologies have significantly constrained their scalability.

To address these challenges, planar chiral metasurfaces have emerged as promising alternatives due to their simplified fabrication and enhanced easy-integration with on-chip systems [21, 22, 23]. Despite these advantages, achieving robust and tunable CD responses in planar configurations remains challenging. Designs based on magnetic dipole resonances [24, 25] generally yield weak CD signals with limited tunability, while those employing in-plane component $T_x$ and $T_y$ of toroidal dipole (TD) resonances [26, 27] suffer from low quality (Q) factors and broad resonance linewidths caused by substantial radiative losses. Moreover, many planar metasurfaces exhibit only moderate chiral selectivity over broad spectral ranges [28, 29], making them unsuitable for narrowband applications that require precise frequency control.

Recent advances in photonics have highlighted bound states in the continuum (BIC) as a powerful mechanism for suppressing radiative losses [30] and achieving exceptionally high Q factors [31, 32]. BICs arise when specific resonant modes remain perfectly confined despite residing within the radiation continuum, thus offering new opportunities for precise light-matter interaction control. Chiral BICs—which combine intrinsic chirality with the BIC mechanism—have further enabled the development of metasurfaces with enhanced circular polarization sensitivity [33] and improved CD responses [34]. However, existing chiral metasurfaces based on BICs often depend on complex geometries or require stringent excitation conditions, heavily limiting their scalability, customizability, and practical applications[35, 36, 37]. Additionally, many of these designs present broad resonance widths [38], undermining their utility in applications necessitating narrowband spectral selectivity and pronounced chiroptical responses.

In this paper, we demonstrate a truly planar chiral metasurface that achieves a giant intrinsic chiral response by utilizing a chiral quasi-bound state in the continuum (Q-BIC) dominated by $T_z$. Our design introduces an additional degree of freedom that enables rapid CD switching within an ultra-narrow frequency range (~0.2 GHz) while maintaining high Q-factors in both simulations (~1,000) and experiments (~73). Furthermore, the metasurface exhibits extrinsic chirality under small-angle oblique incidences (±4°), offering a versatile platform for polarization-sensitive photonic devices and chiral quantum technologies.

## 2. Results and Discussion

### 2.1 Design of planar chiral Q-BIC metasurface

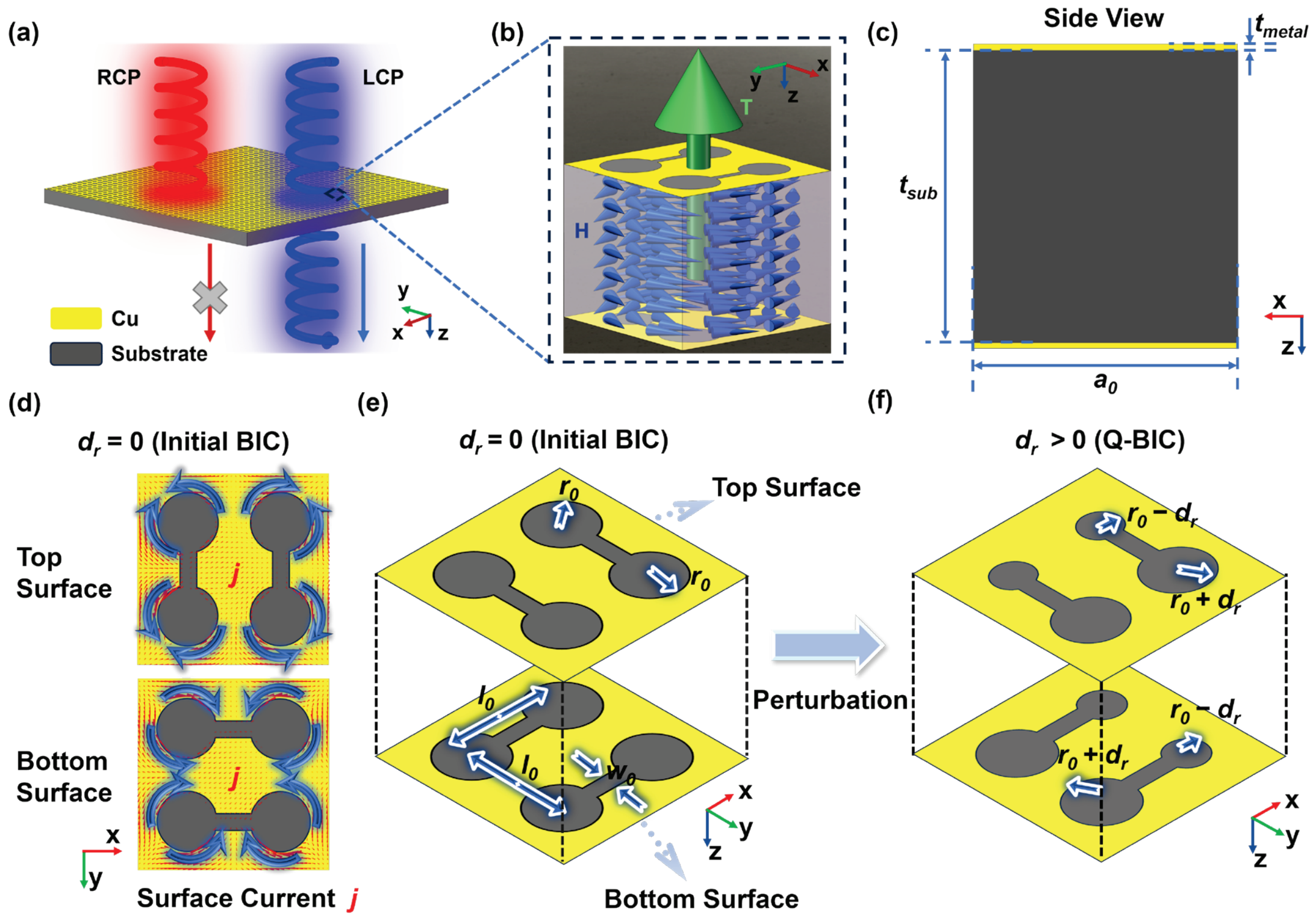

**Fig. 1 | Schematic illustration of planar chiral metasurface supporting toroidal BIC mode**. (a) Illustration of the right-circular polarized (RCP) and left-circular polarized (LCP) electromagnetic (EM) waves incidences on the planar metasurface and the chiral interaction. (b) An enlarged view of a single chiral unit cell highlights the excitation of a TD (**T**) oriented along the *z*-axis (indicated by a large green arrow) and the local circulating magnetic field (**H** field, shown as smaller blue arrows). (c) Side view of the chiral unit cell, highlighting the key parameters: the lattice constant ($a_0$ = 8 mm), substrate thickness ($t_{sub}$ = 0.5 mm), and metal layer thickness ($t_{metal}$ = 18 μm). (d) The surface current distributions $\boldsymbol{j}$ in the *x*-*y* plane for both top and bottom surfaces of the achiral unit cell, which supports the initial BIC mode. (e) Initial BIC configuration with symmetric patterns with its geometric parameters (asymmetry parameter $d_r$ = 0 mm, the width of the narrow stripe $w_0$ = 0.4 mm, circular shape's radius $r_0$ = 1.2 mm, and the center-to-center distance between two circular patterns $l_0$ = 4 mm), preserving $S_4$ symmetry. (f) A perturbed quasi-BIC (Q-BIC) configuration with asymmetry was introduced ($d_r$ = 0.1 mm) on the top and bottom surfaces, allowing for coupling with free-space radiation and resulting in a transition from BIC to Q-BIC mode.

Chiral metasurfaces have attracted significant attention for their ability to manipulate circularly polarized waves and support Q-BIC modes. In this work, we propose a planar chiral metasurface that achieves a pronounced intrinsic chiral response through asymmetric perturbations [39, 40, 41] in a double-layered microstructure [42]. The structure consists of a periodic array of copper patterns on a dielectric substrate (see **Supplementary Materials Note 1** for details on the substrate material), as illustrated in Fig. 1a. The design enables the near-perfect transmission effect of LCP EM waves while significantly suppressing the transmission of the RCP waves, thus realizing a highly selective polarization response that underpins the observed chiroptical effects. An enlarged view of the unit cell of the metasurface is shown in Fig. 1b. The distribution within the unit cell exhibits a circular magnetic field (**H**, indicated by the blue arrows) akin to a magnetic dipole moment. According to the right-hand rule, this leads to a TD oriented along the $z$-axis (**T**, represented by a more prominent green arrow). These features highlight the capability of the metasurface to generate strong chiral responses. The geometric parameters of the metasurface, including the lattice constant, substrate thickness, and metal layer thickness, are shown in Fig. 1c and adopt a vertically stacked configuration with metallic patterns on the top and bottom surfaces of the dielectric substrate. In its initial BIC mode, Fig. 1d illustrates the surface current $\boldsymbol{j}$ (indicated by red arrows) distribution on the top and bottom surfaces of the unit cell in the $x$–$y$ plane. It is clearly observable that the surface current distribution and the direction of current flow (depicted by blue curved arrows) on the top surface exhibit a fully mutually canceling pattern, which forms a loop that reflects a symmetric current distribution around the four holes of the dumbbell-shaped pattern. While on the bottom surface, they are identical to the top surface, but mirror-symmetrical. Therefore, the symmetry of the current around the center of the structure indicates that the LCP and RCP components of the current have reached equilibrium. This configuration is a hallmark of the BIC mode, where energy is confined within the metasurface structure, leading to negligible radiation and no chiral response. Displayed in Fig. 1e, the metasurface exhibits a symmetry-protected BIC when the radii of the top-layer circular elements remain unperturbed ($d_r$ = 0 mm). This symmetric configuration preserves the $S_4$ symmetry of the structure, ensuring non-radiation loss and supporting high-Q resonances. The structural parameters, such as the initial radius ($r_0$), length ($l_0$), and width ($w_0$) of the circular elements of the dumbbell-shaped patterns, further determine the resonance characteristics of the metasurface.

To transition from a BIC to a Q-BIC mode, an asymmetric perturbation ($d_r \neq 0$ mm) is introduced by modulating the radii of the circular elements of the top and bottom layers, as shown in Fig. 1e. This perturbation breaks the $S_4$ symmetry of the structure, enabling coupling between the resonant mode and free-space radiation, which leads to the emergence of the Q-BIC mode. We note that according to our perturbation on the top and bottom layers, the opposite asymmetry parameters $d_r$ (e.g., $d_r = \pm 0.1$ mm) will generate unit cells related by a 180° out-of-plane rotation, thus leading to identical chiroptical responses. In contrast, inverting the sign of $d_r$ only in the top layer, while keeping the perturbation on the bottom layer unchanged, will result in a reversed CD spectrum that creates an enantiomeric pair, as confirmed in our simulations (**Supplementary Materials, Figure S14**). In general, this metasurface design not only provides a tunable platform for planar chirality but also achieves high-performance optical responses by leveraging the Q-BIC mode. The proposed structure effectively bypasses the challenges associated with 3D chiral microstructures, offering significant potential for applications in polarization-sensitive optical devices, chiral sensing, and photonic communication systems.

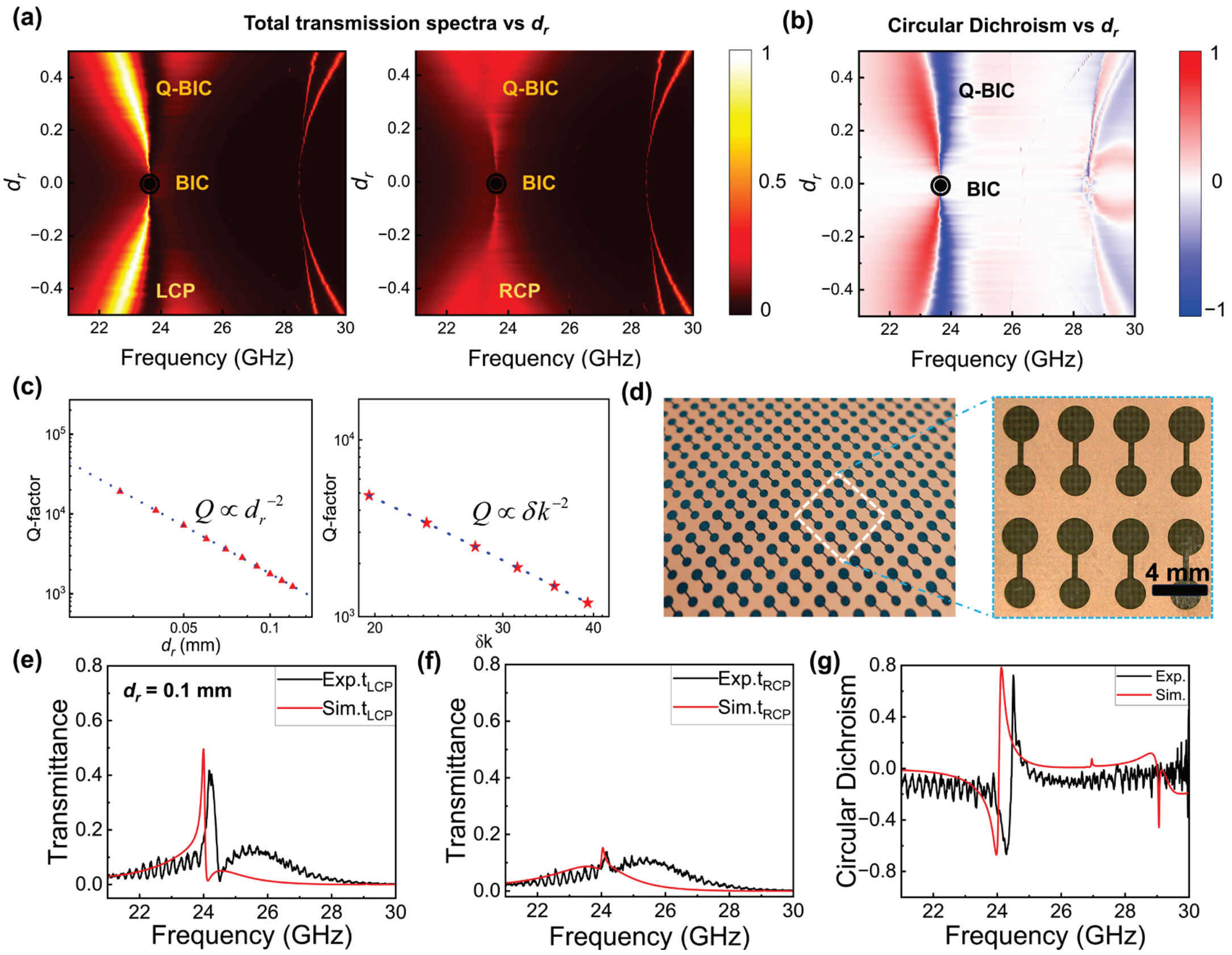

**Fig. 2 | Tunable chiral responses of the Q-BIC metasurface**. (a) The simulated transmission spectra for LCP and RCP incidences as functions of frequency and $d_r$. (b) The corresponding CD spectra as functions of frequency and asymmetry parameter $d_r$, highlighting a sharp resonance near 24 GHz corresponding to the BIC mode (indicated by a small black dot). As $d_r$ increases, the system transitions from a BIC to a Q-BIC mode, resulting in a strong chiral response with a pronounced CD peak. (c) Dependence of the Q factor on $d_r$ and wave number deviation $\delta k$ ($\delta k = \sqrt{k_x^2 + k_y^2}$, the distance away from the Γ point in the reciprocal space); the solid dot lines represent inverse quadratic fittings (log-log scale). (d) Photographs of the fabricated metasurface (middle right), including a zoomed-in view of its periodic unit cells. The radius of the circles of the dumbbells exhibits a difference corresponding to the asymmetry parameter $d_r$ = 0.1 mm. (e-g) Comparison of experimental and simulated data, with transmission spectra for LCP (e) and RCP (f) incidences, and the corresponding CD spectrum shown in (g).

## 2.2 Characterization of intrinsic circular dichroism

Here, we characterize the CD as the transmission difference between LCP and RCP incidences by Eq. (1), while the cross-polarized and co-polarized transmission coefficients are derived from the Jones matrix formalism by Eq. (2) [43]:

$$CD=\frac{t_{RCP}-t_{LCP}}{t_{RCP}+t_{LCP}}=\frac{\left(\left(t_{LR}\right)^2+\left(t_{RR}\right)^2\right)-\left(\left(t_{RL}\right)^2+\left(t_{LL}\right)^2\right)}{\left(\left(t_{LR}\right)^2+\left(t_{RR}\right)^2\right)+\left(\left(t_{RL}\right)^2+\left(t_{LL}\right)^2\right)}, \tag{1}$$

$$\begin{pmatrix} t_{RR} & t_{RL} \\ t_{LR} & t_{LL} \end{pmatrix}=\frac{1}{2}\begin{pmatrix} t_{xx}-t_{yy}+i\left(t_{xy}+t_{yx}\right) & t_{xx}+t_{yy}-i\left(t_{xy}-t_{yx}\right) \\ t_{xx}+t_{yy}+i\left(t_{xy}-t_{yx}\right) & t_{xx}-t_{yy}-i\left(t_{xy}+t_{yx}\right) \end{pmatrix}, \tag{2}$$

where $T_{ij} = |t_{ij}|^2$ ($i, j = R, L$, or $x, y$; with $R$ and $L$ denoting RCP and LCP components, while $x$ and $y$ denoting $x$- and $y$-linearly polarized components) represents the transmissions. Here, we adopt the convention that $i$ denotes the output polarization and $j$ denotes the input polarization for all the elements of the Jones matrix $t_{ij}$. We also stipulate that the LCP (RCP) component corresponds to $(E_x \pm iE_y)/\sqrt{2}$ in our investigation as we adopt the $e^{+i\omega t}$ time-harmonic convention to facilitate comparison with experiments. Moreover, we define $t_{\mathrm{LCP}}$ ($t_{\mathrm{RCP}}$) as the total transmittance under LCP (RCP) incidences. Taking $t_{\mathrm{LCP}}$ as an example, it includes both co-polarized (LCP→LCP) transmittance coefficient $t_{\mathrm{LL}}$ and cross-polarized (LCP→RCP) transmittance coefficient $t_{\mathrm{RL}}$, that is, $t_{\mathrm{LCP}} = |t_{\mathrm{LL}}|^2 + |t_{\mathrm{RL}}|^2$, and similarly for RCP, $t_{\mathrm{RCP}} = |t_{\mathrm{RR}}|^2 + |t_{\mathrm{LR}}|^2$. This definition aligns with the convention in circular dichroism measurements, where the total transmitted intensity under each handedness is considered, independent of the output polarizations.

Fig. 2a depicts the total simulated transmission spectra for LCP and RCP at normal incidence as a color map of frequency and $d_r$, in which a near-perfect transmission peak and a near-zero transmission dip corresponding to the Q-BIC mode under LCP incidence are observed. On the other hand, although the RCP spectrum under $d_r = 0$ mm exhibits near-zero linewidth, it remains nearly unchanged under $d_r \neq 0$ mm, reflecting the strong polarization selectivity of the Q-BIC mode. This selective coupling demonstrates the proposed metasurface's ability to control circularly polarized light by leveraging symmetry-breaking perturbations. Based on the total transmission spectra of LCP and RCP of Fig. 2a, in Fig. 2b, the CD spectrum exhibits a near-zero linewidth phenomenon at approximately 24 GHz, indicated by a black dot, when the metasurface preserves $S_4$ symmetry as $d_r = 0$ mm, which is owing to a symmetry-protected BIC mode at the Γ point that does not couple to the far field input at normal incidence. As the $d_r$ varies, which breaks the $S_4$ symmetry, the BIC evolves into a Q-BIC mode. As a result, the linewidth of the CD spectrum broadens, and a strong chiral response emerges. Notably, the CD reaches a maximum value of +0.97 at around 24 GHz when $d_r = 0.07$ mm, with the CD scale varying from approximately −0.97 (blue region) to +0.97 (red region). In general, the peak CD values and their frequencies depend on the specific $d_r$ value yet remain giant (peak CD > +0.7 or <−0.7) for non-zero $d_r$ ranging between −0.5 mm and +0.5 mm, as clearly shown in Fig. 2b. To validate support of the Q-BIC in proposed chiral metasurface, we calculate and fit the relationships between its Q factor (see **Supplementary Materials Figure S1** for details of Q-factors under different asymmetry $d_r$) and the asymmetry parameter $d_r$, as well as the wave number δk deviated from the Γ point, as illustrated in Fig. 2c. The Q factor exhibits expected inverse square relationships with these two parameters, consistent with the theoretical predictions for a first-order BIC [44]. The relationship between the Q factor and the asymmetry parameter $d_r$ also provides a convenient method for tuning the Q factor of chiral responses. To thoroughly assess the performance of our tunable Q-BIC chiral metasurface, we calculate the CD and plot the spectra for various thicknesses of substrate $t_{sub,}$ (see **Supplementary Materials Figure S2** for detailed results) using COMSOL Multiphysics (see **Supplementary Materials Note 2** for physical settings in COMSOL Multiphysics).

To further verify the giant intrinsic chirality of our metasurface, we conduct experimental characterization on the sample with asymmetry parameter $d_r$ = 0.1 mm (see **Supplementary Materials Figure S3** for details of experimental setup). The reason we chose to manufacture tested samples with asymmetry parameter $d_r$ = 0.1 mm rather than 0.07 mm is that the typical manufacturing tolerance and standard minimum feature size for PCB fabrication is 0.1 mm. Fig. 2d presents photographs of the samples fabricated using the standard printed circuit board (PCB) technology for $d_r$ = 0.1 mm, which supports the Q-BIC mode. The magnified inset clearly shows that the metallic surface of the prepared metasurface samples exhibits distinct and uniform dumbbell-shaped periodic patterns. Figs. 2e-g illustrate the total transmission spectra $t_{\mathrm{LCP}}$ and $t_{\mathrm{RCP}}$, as well as the CD spectra, derived from simulations and measurements based on the transmission Jones matrix (see **Supplementary Materials Note 3** for the detailed definition and calculation of the Jones matrix) for the fabricated metasurfaces. The measured spectra are in excellent agreement with the simulated results, validating the chiroptical performance of the metasurface. As illustrated in Fig. 2e, the proposed chiral metasurface exhibits a strong response when subjected to a normal LCP incident wave at a frequency of 24 GHz, where a sharp Fano transmission peak with a transmittance approaching 0.5 can be observed in the $t_{LCP}$ transmission spectrum. Conversely, the transmittance for the RCP incident wave, as shown in Fig. 2f, is nearly zero. Following the previously established equation for CD, we present simulated and experimentally measured CD spectrum in Fig. 2g, revealing a pronounced peak value of up to 0.8 at 24 GHz, along with a notable CD switching phenomenon occurring within a very narrow frequency range of approximately 0.2 GHz, whose narrow linewidth leads to a steep spectral slope in the CD, switching from positive to negative values. This phenomenon originates from the polarization-selective asymmetric radiation of the toroidal Q-BIC, which we will discuss below. Moreover, as shown in **Supplementary Materials Figure S2**, there exists an optimal substrate thickness that maximizes the CD response. In brief, the substrate thickness significantly influences the near-field coupling and phase conditions that govern the toroidal Q-BIC excitation, thereby affecting the CD strength. In contrast, the metal thickness has a negligible impact due to its sufficient thickness relative to the skin depth. It is also important to emphasize that the observed transmission and CD responses do not originate from background leakage through the cylindrical holes in the copper layers. At the resonant frequency of ~24 GHz, the fundamental cut-off frequencies of these partially filled metallic holes, whose diameters are 2.6 mm and 2.2 mm, are approximately 42.3 GHz and 50 GHz, respectively, well above our spectral window (see **Supplementary Materials Note 11**). Therefore, no propagating mode can be supported inside the holes, and the measured transmission must arise solely from the symmetry-broken toroidal Q-BIC resonance.

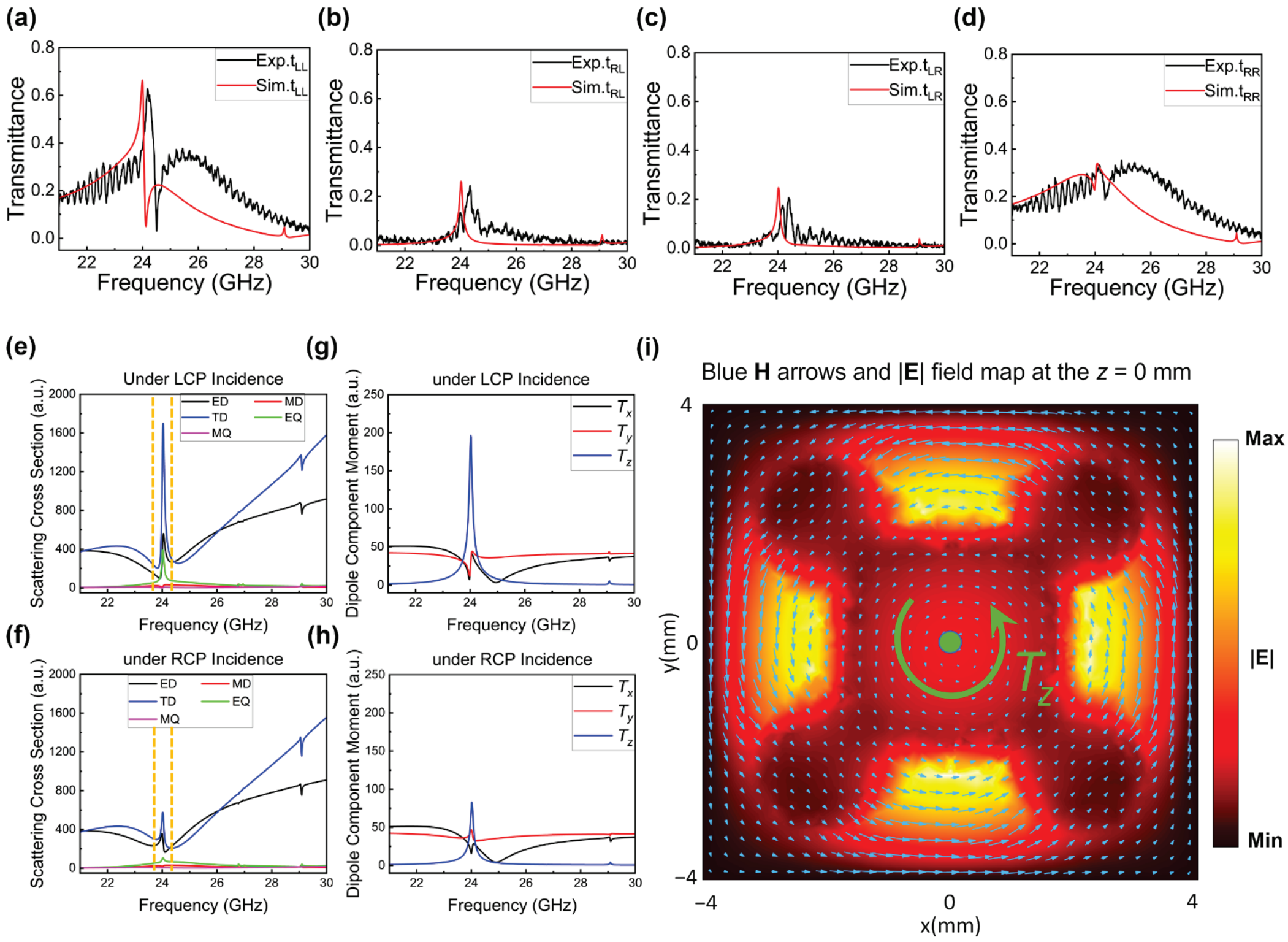

**Fig. 3 | Multipole contributions for the intrinsic chirality of the Q-BIC mode**. (a-d) Simulated and experimental transmission spectra for circular polarizations, showing co-polarized ($t_{LL}$ (a) and $t_{RR}$ (d)) and cross-polarized ($t_{RL}$ (b) and $t_{LR}$ (c)) transmission coefficients. (e,f) Calculated scattering cross sections of several first orders of multipoles, including electric dipole (ED), magnetic dipole (MD), TD, electric quadrupole (EQ), and magnetic quadrupole (MQ) under LCP (e) and RCP (f) incidences, respectively. (g, h) Amplitudes of the components ($T_x$, $T_y$, and $T_z$) for LCP (g) and RCP (h) incidences. (i) Simulated normalized in-plane electric field |**E**| distribution on the x-y cut plane ($z$ = 0 mm), overlaid with magnetic field vectors (**H**, blue arrows) and the corresponding component vector $T_z$ indicated by circular magnetic field arrows.

### 2.3 Multipolar analysis and characterization of TD

To explore the mechanisms underlying the chiroptical responses of the Q-BIC metasurface, we analyze the transmission spectra, multipolar scattering contributions, and eigenmode near-field distributions under LCP and RCP wave incidences. Fig. 3a-d displays both simulated and experimental transmission spectra. The co-polarized ($t_{LL}$, $t_{RR}$) and cross-polarized ($t_{LR}$, $t_{RL}$) transmission coefficients reveal pronounced dips corresponding to the Q-BIC resonance (see **Supplementary Materials Figure S4** for linear transmission spectra ($t_{xx}$, $t_{yx}$, $t_{xy}$, $t_{yy}$) under the asymmetry parameter $d_r$ = 0.1 mm of the chiral Q-BIC metasurface), whose position is sensitively dependent on the incident polarization. Despite minor discrepancies, the experimental data agree well with simulations, confirming the ability of the metasurface to achieve highly selective polarization-dependent transmission. The observed dips in the transmission spectra correspond to the Q-BIC resonance, which is sensitive to incident polarization. According to the scattering theory [45], the scattering cross sections of this mode, decomposed into lower-order multipoles, are displayed in Fig. 3e and Fig. 3f for LCP and RCP incidences, respectively (see **Supplementary Materials Note 4** for definitions of these dipole moments and scattering cross sections). TD dominates the scattering process, contributing significantly to the chiral response. Other multipoles, including the ED, MD, EQ, and MQ, also contribute to the overall optical response but are secondary compared to the TD moment.

Fig. 3g and Fig. 3h show the amplitude of the TD components ($T_x$, $T_y$, and $T_z$) under LCP and RCP incidences. The $T_z$ is strongly excited under LCP incidence near the Q-BIC resonance frequency, while its excitation is suppressed under RCP incidence. This polarization-dependent behaviour emphasizes the critical role of the $T_z$ in enabling the chiral properties of the metasurface. Finally, Fig. 3i illustrates the simulated in-plane electric field (|**E**|) distributions of the Q-BIC mode at the *x*-*y* cutting plane where the *z* position is 0 mm, that is, the mid-plane of the metasurface (see **Supplementary Materials Figure S5** for the electric field distribution amplitude (|**E**|) of the Top and Bottom Surfaces of unit cell of the metasurface). The field patterns reveal a characteristic circulating distribution, indicative of the TD resonance. Overlaid magnetic field vectors (**H**, blue arrows) and associated magnetic vectors (white arrows) further confirm the presence of the TD and other multipolar contributions. We also calculate the *reactive helicity density* (RHD) and *optical chirality density* (OCD) field maps to demonstrate the chiral density shown in the Q-BIC chiral metasurface (see **Supplementary Materials Figure S6** for details of the eigenfrequency field maps of OCD and RHD) [46]. These results provide a clear visualization of the near-field characteristics underlying the chiral response of the metasurface, further confirming that the observed CD originates from the excitation of the $T_z$. The $S_4$ symmetry of the metasurface when $d_r$ = 0 mm efficiently suppresses in-plane electric and magnetic dipole moments, enabling a dominant and isolated out-of-plane toroidal moment in the BIC case. When the $S_4$ symmetry is slightly broken, the $T_z$ moment still dominates the Q-BIC mode, and its interplay with the in-plane electric and toroidal moments leads to a strong chiral response (see **Supplementary Materials Figure S11** for multipole analysis of BIC and Q-BIC modes).

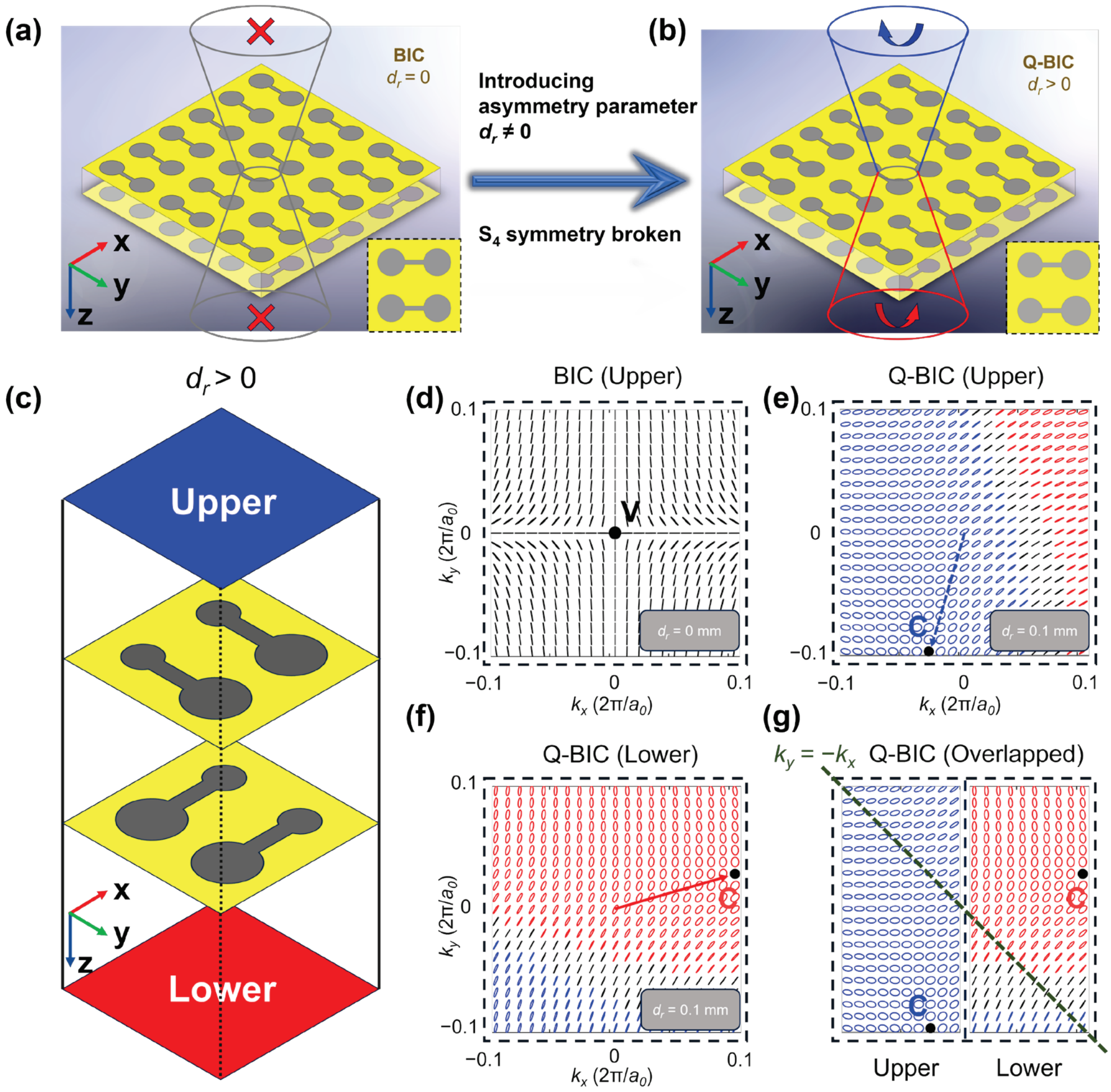

**Fig. 4 | Evolution of circular polarization from BIC to Q-BIC modes**. (a) Schematic of the metasurface structure showing the system state in the BIC mode, with a zero-asymmetry parameter ($d_r$ = 0 mm), maintaining perfect $S_4$ symmetry. (b) Q-BIC mode induced by $d_r > 0$, breaking symmetry to generate distinct LCP/RCP radiation. (c) The Q-BIC unit cell with blue/red-colored Upper/Lower surfaces (40 mm away from the metasurface) for characterizing polarization states of the far-field radiation. (d-f) Evolution of C points over momentum space for the metasurfaces of different asymmetry parameters $d_r$. The elliptical polarizations are represented by ellipses of blue/red-colored corresponding to left-/right-handed states, whereas the black lines represent linear polarizations: (d) The eigen-polarization map of the BIC mode under $d_r$ = 0 mm, a polarization singularity (a V point with topological charge $S = -1$) is observed at the Γ point in the upper surface of the far-field radiation, (e-f) For $d_r$ = 0.1 mm of Q-BIC mode, the at-Γ V point splits into a pair of LCP/RCP C points of the far-field radiation: (e) LCP C point with $-\frac{1}{2}$ charge located in Upper surface of the far-field radiation, (f) RCP C point with $-\frac{1}{2}$ charge located in Lower surface of the far-field radiation. (g) Polarization map of the left half of the Upper surface overlapping the right half of the Lower face in Q-BIC mode, showing the LCP/RCP C points symmetrically distributed along the $k_y = -k_x$ straight line.

## 2.4 The evolution of BIC polarization singularity

The metasurface design exploits symmetry breaking to transition from a symmetry-protected BIC mode to a Q-BIC mode, facilitating the emergence of distinct circular polarization states. This transition mechanism is schematically shown in Fig. 4a-b. As is shown in Fig. 4a, when the asymmetry parameter ($d_r$ = 0), the metasurface structure retains perfect $S_4$ symmetry, corresponding to the BIC mode with negligible radiation loss. Upon introducing a non-zero asymmetry ($d_r$ > 0), symmetry breaking occurs, and the system evolves into a Q-BIC mode, supporting chiral Q-BIC responses. To further explore the evolution of polarization singularities, we plot the polarization maps in momentum space for both BIC and Q-BIC modes (corresponding to the $TM_4$ mode in Fig. 5b), with varying asymmetry parameter $d_r$ values (shown in Fig. 4d-g). In the absence of perturbation ($d_r$ = 0), the far-field radiation on the Upper surface (see **Supplementary Materials Figure S7** for details of far-field radiation on the Upper and Lower surfaces with various asymmetry $d_r$ values) above the metasurface exhibits a V point singularity (represented by a black dot) at the Γ point with topological charge $S = -1$ (see **Supplementary Materials Note 5** for definitions and calculation formula of topological charge $S$), indicating polarization singularity of the BIC mode (Fig. 4d). As the asymmetry parameter is perturbed ($d_r$ = 0.1 mm), the at-Γ V point splits into two symmetrically distributed C points away from Γ point (see **Supplementary Materials Figure S8** for details of far-field electric field patterns of these two C points that represent unidirectional circular-polarized radiation).

In fact, as shown in Fig. 4e-f: the LCP C point with topological charge $S = -\frac{1}{2}$ is located in the Upper surface slice of the far-field radiation shown in Fig. 4e, while the RCP C point with topological charge $S = -\frac{1}{2}$ is in the Lower surface slice of the far-field radiation shown in Fig. 4f. Furthermore, the overlapped polarization map that combines the Upper and Lower surface slices of the far-field radiation (shown in Fig. 4g) reveals the evolution of the at-Γ V point into a pair of LCP/RCP C points, both of which are star C points [47] and mirror-symmetrically located around the Γ point relative to the $k_y = -k_x$ line. This results in the formation of a characteristic polarization distribution around Γ point with opposite circular polarizations on the Upper and Lower surface slices of the far-field radiation, thus achieving a significant chiral response [48], both intrinsic (at-Γ) and extrinsic (off-Γ). In general, this evolution underscores the robustness of the topologically distinct polarization states, and driven by symmetry breaking, they shape the chiral responses of the metasurface.

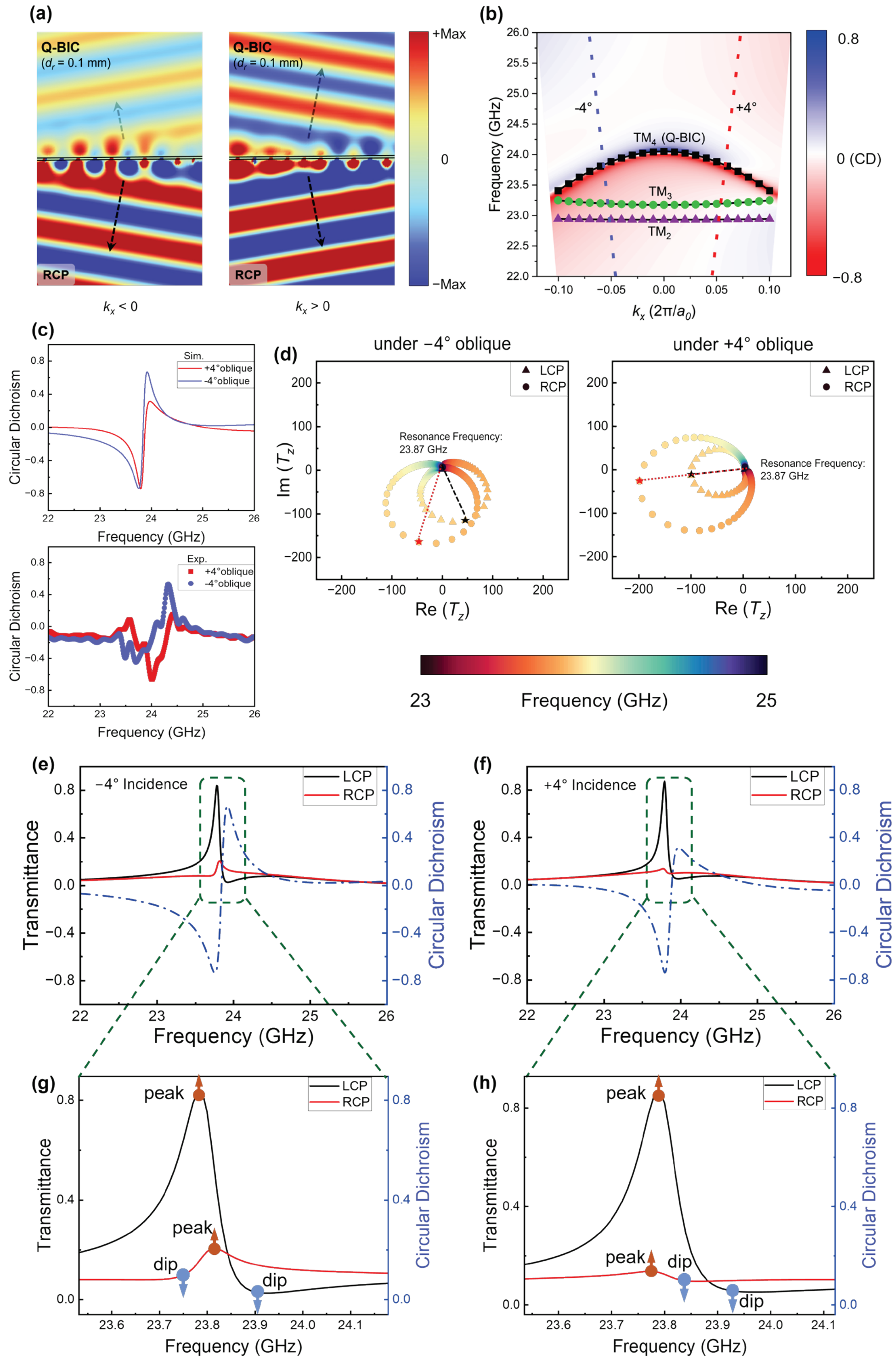

**Fig. 5 | Asymmetric CD response of Q-BIC metasurface under small-angle incidence**. (a) The circularly-polarized (RCP) electric field patterns for off-Γ point ($k_x \neq 0$) with $d_r$ = 0.1 mm, showing asymmetric radiation patterns. Here, we consider $k_x = \pm 0.2\pi/a_0$ as a specific example. (b) Band structures of the metasurface highlight the CD and three typical transverse magnetic (TM)-like modes (TM$_2$, TM$_3$, TM$_4$) as functions of $k_x$ and frequency. (c) The simulated and measured CD spectra under oblique incidence conditions of ±4°. (d) The complex-plane trajectories of excited TD component $T_z$ under LCP and RCP ±4° oblique incidences, as $|T_z|$ =

$\sqrt{\mathrm{Re}(T_z)^2 + \mathrm{Im}(T_z)^2}$, which reflect the excitation strength of the Q-BIC mode. The resonance frequency points are marked by black stars for LCP, and by red stars for RCP, and also indicated by the dashed lines. Thus, the trajectories for the −4° and +4° angles reveal excitations of the toroidal Q-BIC mode with starkly different phases under opposite angles of incidence. Each trajectory forms a closed loop due to the narrow linewidth of the Q-BIC mode. The quantities $\mathrm{Re}(T_z)$ and $\mathrm{Im}(T_z)$ are in arbitrary units. (e-f) Transmission spectra of LCP (black solid line) and RCP (red solid line) together with the corresponding CD (blue dashed line) spectra under −4° (e) and +4° (f) oblique incidences, respectively. (g-h) The magnified views of the spectra in the 23.6-24.1 GHz range, with peak and dip symbols marked for the Fano-like spectra. Under −4° incidence, the LCP transmission spectrum shows a peak-dip sequence (orange→blue arrow markers), while the RCP transmission spectrum exhibits the inversed dip-peak (blue→orange arrow markers) sequence due to the different phases of the excited toroidal Q-BIC, which results in the rapid CD switching. Under +4° incidence, both the LCP and RCP transmission spectra display the peak-dip sequence due to the identical phases of the excited toroidal Q-BIC, which results in the absence of rapid CD switching.

## 2.5 Asymmetric far-field responses and $T_z$ switching dynamics

To further analyze the chiral optical response of the metasurface supporting Q-BIC for oblique incidence, we investigate the radiation of eigenmodes at off-Γ points. As an example, Fig. 5a illustrates the RCP electric field maps for the Q-BIC eigenmode with $k_x \neq 0$ and an asymmetry parameter of $d_r$ = 0.1 mm. More specifically, we consider $k_x = \pm 0.2\pi/a_0$, and under both cases, the RCP field of the eigenmode predominantly radiates downward (indicated by dashed arrows denoting the oblique radiation directions), with much weaker radiation upward (indicated by semi-transparent dashed arrows). However, due to the breaking of out-of-plane mirror symmetries when $d_r \neq 0$, the radiation upward is also strikingly different for opposite values of $k_x$. This asymmetric radiation behaviour underpins the strong but different chiral responses of the metasurface for opposite angles of incidence, which originates from the interplay between the in-plane and out-of-plane dipole moments of the toroidal Q-BIC mode.

For quantitative analysis, Fig. 5b illustrates the detailed relationship between the CD and band structure of the metasurface with respect to the wave vector $k_x$, and frequency. Three TM-like bands, namely $TM_2$, $TM_3$, and $TM_4$, are identified and represented by magenta triangles, green circles, and black squares (see **Supplementary Materials Figure S9** for details of the band structure of $TM_1$ mode). The $TM_4$ band at $k_x = 0$ corresponds to the Q-BIC mode at 24.0 GHz, giving rises to a near-unity intrinsic CD value, and a chiral near-flat band [49]. While the $TM_2$ and $TM_3$ modes also maintain a flat band structure, the metasurface does not exhibit a strong chiral response at their respective frequencies. However, the interplay of the $TM_2$, $TM_3$, and $TM_4$ modes facilitates the polarization-selective behaviour of the metasurface. Both of the simulated and measured CD spectra under oblique incidence with opposite incident angles of ±4° are shown in Fig. 5c. The results confirm that the metasurface retains its strong chiral properties (CD > 0.7) even when the incidence angle deviates slightly from normal incidence. A more interesting point is the asymmetric responses even under such small incident angles. To clarify how these asymmetry responses arise from the Q-BIC mode dominated by $T_z$ component, Fig. 5d presents the complex-plane trajectories of the $T_z$ under LCP and RCP ±4° oblique incidences. The trajectories are plotted as (Re($T_z$), Im($T_z$)) with $|T_z| = \sqrt{\mathrm{Re}(T_z)^2 + \mathrm{Im}(T_z)^2}$, showing a clear crossing behaviour for $|T_z|$ amplitude under −4° oblique incidence and no such crossing for +4° incidence. We also calculate the TD's real and imaginary parts of all its three components ($T_x$, $T_y$, $T_z$) under LCP and RCP incidences under oblique incident angles of ±4° (see **Supplementary Materials Figure S10** for details). In fact, under such small angles of incidence, the transmission still mainly comes from the in-plane multipolar components ($T_x$, $T_y$, $p_x$, $p_y$), while the out-of-plane component $T_z$ indicates of the excitation strength of the Q-BIC mode (see **Supplementary Materials Figure S11** for details). Moreover, the ratios between the in-plane multipolar components and the out-of-plane component $T_z$ under opposite angles of incidence are generally similar (see **Supplementary Materials Note 12**), while the trajectories in Fig. 5d highlight the starkly different excitations of the toroidal Q-BIC mode under such opposite angles of incidence. The trajectory crossing for the −4° oblique incidence angle is closely tied to the phase differences in $T_z$ under LCP and RCP incidences. Phase analysis confirms that $T_z$ exhibits clear phase detuning under −4° incidence but not +4° incidence, directly governing the interference asymmetry (see **Supplementary Materials Note 13 and Figure S16**). Moreover, phase detuning also manifests at other positive and negative oblique incident angles. In general, the phase detuning persists systematically over all negative incident angles (−2° to −6°), while being absent at positive angles, thereby confirming the systematic nature of this mechanism (see **Supplementary Materials Note 14 and Figure S17**). Specifically, the circularly polarized transmissions at ±4° incidences display Fano-like spectra [50, 51] with adjacent peak and dip features, as illustrated in Figs. 5(e) and 5(f). We also conduct the transmittance analysis from 0° to ±6°, which reveals that negative angles produce opposite Fano asymmetries for LCP/RCP transmission spectra, while positive angles yield identical asymmetries (see **Supplementary Materials Note 15 and Figure S18**). Near the resonance frequency, the spectra, magnified in Figs. 5(g) and 5(h), show a reversed dip-peak or peak-dip sequence for −4° oblique incidence due to phase differences between LCP and RCP transmission spectra, but no such reversal occurs for +4° oblique incidence. These findings elucidate the rapid switching of CD at zero or negative incidence angles, which diminishes quickly at positive angles (see **Supplementary Materials Note 16 and Figure S19**). In general, these spectra for LCP and RCP incidences further elucidate the toroidal characteristics of the chiroptical responses of the metasurface. Our results highlight the capability of our metasurface to differentiate small oblique angles of incidence, which is crucial for applications such as chiral sensing and polarization control.

## 3. Conclusion

In this work, we have experimentally demonstrated the giant intrinsic (under normal incidence) chiroptical responses in a truly planar and tunable chiral metasurface supporting a toroidal Q-BIC mode. By selectively exciting resonances dominated by the $T_z$ component of TD, we identify a pronounced chiral response with near-unity CD values. Moreover, the intrinsic chiral response is marked by a rapid switching phenomenon and a sharp transition from positive to negative CD values within a narrow frequency window of about 0.2 GHz, opening and facilitating feasible possibilities for high-precision and adjustable control over optical communication and sensing technologies. Meanwhile, the metasurface also exhibits highly asymmetrical CD spectra under opposite oblique incidences of small angles (±4°). In fact, the transition from BIC to Q-BIC mode is controlled by symmetry-breaking perturbations, as evidenced by the evolution from the at-Γ V point to a pair of C points in momentum space. These polarization singularities play a critical role in enabling the observed asymmetric chiral responses for oblique incidences. The toroidal Q-BIC metasurface, owing to its near-unity intrinsic chirality, and monolithic and easily-fabricated planar architecture, offers a versatile platform for a wide range of applications, including chiral quantum optics, bio-enantiomer-selective sensing, and polarization-selective photonic devices.

**Acknowledgement**

This research was supported by the National Natural Science Foundation of China (No. 12304348), Guangdong Basic and Applied Basic Research Foundation (No. 2025A1515011470), Guangdong University Featured Innovation Program Project (2024KTSCX036), Guangdong Provincial Project (2023QN10X059), Guangzhou-HKUST(GZ) Joint Funding Program (2025A03J3783), Guangzhou Municipal Science and Technology Project (2024A04J4351), Guangzhou Young Doctoral Startup Funding (2024312028), and Guangzhou Higher Education Teaching Quality and Teaching Reform Engineering Project (2024YBJG087).